\documentclass[aps,prl,twocolumn,groupedaddress,amsmath,amssymb]{revtex4-1}
\usepackage{mathtools}
\usepackage{graphicx,caption,subcaption}  
\usepackage{dcolumn}   
\usepackage{bm}        
\usepackage{verbatim}   
\usepackage{wrapfig}
\usepackage{float, array}
\usepackage{amsmath}
 \usepackage{amssymb}
\usepackage{verbatim, url}
\usepackage{fullpage}
\usepackage{enumerate}
\usepackage{bm}
\usepackage{notes2bib}
\usepackage{lpic}
\usepackage{pgfplots}
\usepackage[T1]{fontenc}
\usepackage[latin1]{inputenc}
\usepackage[extra]{tipa}
\usepackage{accents}
\usepackage{epstopdf}

 \usepackage{amsthm}

\newlength{\dhatheight}

\newcommand{\bra}[1]{\langle #1|}
\newcommand{\ket}[1]{|#1\rangle}

\newcommand{\nn}{\nonumber}
\newcommand{\f}[2] {\frac{#1}{#2}}

\newcommand{\m}[1]{\mathcal{#1}}

\newcommand{\beq}{\begin{equation}}
\newcommand{\eeq}{\end{equation}}
\newcommand{\beqn}{\begin{eqnarray}}
\newcommand{\eeqn}{\end{eqnarray}}

\DeclarePairedDelimiter\abs{\lvert}{\rvert}%
\DeclarePairedDelimiter\norm{\lVert}{\rVert}%

\makeatletter
\let\oldabs\abs
\def\abs{\@ifstar{\oldabs}{\oldabs*}}
\let\oldnorm\norm
\def\norm{\@ifstar{\oldnorm}{\oldnorm*}}
\makeatother

\begin{document}

\title{Optomechanical Toy Model for Gravitationally Induced Decoherence: Exact Solution}
\author{Seyyed M.H. Halataei}
\affiliation{
   Department of Physics, University of Illinois at Urbana-Champaign,\\
   1110 West Green St, Urbana, Illinois 61801, USA
   }
   
\date{March 29, 2017}
\begin{abstract}
I present the exact solution of a toy model for gravitationally induced decoherence. The toy  model has Hamiltonian resembling optomechanical systems. It is an oscillator system coupled through its energy to an oscillator heat bath. I find the decoherence effect of vacuum fluctuations at zero temperature. Also for a finite bath I show that the decoherence is in general present and the system does not return to its initial coherence unless the fundamental frequencies of the bath have rational ratios. 
 \end{abstract}

\maketitle
The decoherence effect of gravitational perturbations has been studied by various authors by use of Born-Markov approximation \cite{Blencowe2013}. While the approximation is valid in certain limits it fails for zero temperature \cite{Blencowe2014pc}. In this paper, we solve the toy model proposed by Blencowe as an illustrative example which may shed some lights on the gravity problem\cite{Blencowe2013,Blencowe2014pc}. We solve the problem almost exactly without any use of Born-Markov approximation. 

Since the toy model resembles genuine models of optomechanics \cite{Aspelmeyer2014} and variation of that may be used in quantum optics and cavity QED, we pay extra attention to finite bath. We show that in a finite bath the decoherence still takes place unless the fundamental frequencies of the bath ratios equal to rational numbers. 
 
The toy model of Ref \cite{Blencowe2013} is an oscillator system with frequency $\omega_0$ coupled via its energy to an oscillator heat bath, 
\beq
H = \hbar \omega_0 a^{\dagger} a  + \hbar \omega_0 a^{\dagger} a \sum_i \lambda_i \f{q_i}{\Delta_i}  + \sum_i \left( \f{p_i^2}{2 m_i} + \f{1}{2} m_i \omega_i^2 q_i^2 \right)
\eeq
where $\Delta_i = \sqrt{\hbar/(2 m_i \omega_i)}$ is the ith bath oscillator's zero-point uncertainty.  In following we  denote the Hamiltonian of the system by $H_S = \hbar \omega_0 a^\dagger a$ and for convenience choose units in such a way that $\hbar = 1$. We also write the total Hamiltonian in terms of annihilation and creation operators of the bath oscillators and drop the c-number vacuum energy of the bath, $\sum_i \f{\hbar \omega_i}{2}$, as follows
\beq \label{H}
H = H_S + H_S \sum_i \lambda_i (a^\dagger_i + a_i)   + \sum_i  \omega_i a_i^\dagger a_i
\eeq
As pointed out in \cite{Halataei2017d} the vacuum energy of the bath does not appear in the original construction of the oscillator heat bath model by Caldeira and Leggett (App. C of \cite{Caldeira83}). So it is more accurate not to include that term in the Hamiltonian and the density matrix of the bath in thermal equilibrium. 

The universe $U$ (which is composed of the system plus the bath) is initially decoupled with total density matrix
\beq
\hat{\rho}_U (0) = \hat{\rho}_S (0) \otimes \hat{\rho}_B
\eeq
where $\rho_S(0)$ is the initial density matrix of the system and the bath is in thermal equilibrium with density matrix 
\beq
\hat{\rho}_B = \bigotimes_i \hat{\rho}_{B_i} = \bigotimes_i \f{e^{- \omega_i a_i^\dagger a_i/k_B T}}{Z_i}
\eeq
where the partition function of the ith oscillator of the bath is $Z_i = Tr_{B_i} \{ e^{-\omega_i a_i^\dagger a_i/k_B T} \}$. As the universe evolves by time it entangles the system and the bath. The density matrix of the universe $\rho(t)$ in no longer a direct product of the that of the system and the heat bath. One has to calculate the reduced density matrix $\rho_S(t)$ as follows to obtain information about the system
\beq
\hat{\rho}_S (t) = \text{Tr}_B \hat{\rho}_U (t)
\eeq
The question is how the off diagonal elements of the reduced density matrix of the system $\rho_S(t)$ decay in time ? 

Without resort the Bron-Markov approximation, this problem is solved by Schlosshauer in Ref. \cite{Schlosshauer2007} for a single spin-$1/2$ particle in interaction with an oscillator heat bath. In \cite{Schlosshauer2007} the total Hamiltonian is similar to \eqref{H} and the Hamiltonian of the system  $H_S \propto \sigma_z$ describes a spin-$1/2$ particle in magnetic field along the z axis. 

We follow Schlosshauer scheme below and adapt it for the oscillator system. In addition we introduce a lower cut off frequency for the bath modes and study its effect, as this can appear in a gravitational problem. As for a finite heat bath and the {\it false decoherence} effect we make, however, a different comment than that given by Ref. \cite{Schlosshauer2007} (see below).

The Hamiltonian \eqref{H} consists of three parts, the system Hamiltonian $H_S$, the interaction Hamiltonian $H_{int} = H_S \sum_i \lambda_i (a_i^\dagger + a_i)$ and the bath Hamiltonian $H_B = \sum_i \omega_i a_i^\dagger a_i$,
\beq
H = H_S + H_{int} + H_B.
\eeq
Since the decaying behavior of the reduced density matrix is of interest here (and not its oscillatory one) we go to the interaction picture and solve the problem there. In this picture the interaction Hamiltonian $H_{int}(t)$ evolves as  
\beqn
H_{int} (t) &=& e^{i (H_S + H_B) t} H_{int} e^{- i (H_S + H_B) t} \\
&=& H_S \sum_i \lambda_i (a_i^\dagger e^{i \omega_i t} + a_i e^{-i \omega_i t})
\eeqn
From this one can construct the evolution operator in the interaction picture, 
\beq
\hat{U}(t) = \m{T} e^{-i \int_0^{\infty} dt' H_{int}(t')}
\eeq
where $\m{T}$ denotes the time ordered product. Since the commutator of the interaction Hamiltonian for two different times is a c-number,
\beq
[H_{int} (t), H_{int}(t')] = -2 i \sum_i \lambda_i^2 \sin \omega_i (t-t'),
\eeq
one can write $U(t)$ as 
\beq \label{U}
\hat{U}(t) = e^{i \phi(t)} \exp\left[ - i \int_0^t dt' H_{int}(t') \right] \equiv e^{i \phi(t)} \hat{V}(t)
\eeq
where $\phi(t)$ is a global time dependent phase factor and there is no time ordering above \cite{Schlosshauer2007}. Taking the integral one finds 
\beq \label{V}
\hat{V}(t) = \exp \left[ \f{H_S}{2} \sum_i \left( \kappa_i(t) a_i^\dagger + \kappa_i^*  (t) a_i \right) \right]
\eeq
where
\beq
\kappa_i (t) \equiv 2 \f{\lambda_i}{\omega_i} (1 - e^{i \omega_i t}) 
\eeq
Now one can calculate the evolution of the reduced density matrix $\rho_S(t)$ by use of \eqref{U}-\eqref{V},
\beqn
\hat{\rho}_S(t) &=& \text{Tr}_B \left[ \hat{U}(t) \hat{\rho}_U(0) \hat{U}(t)^{-1} \right] \\
&=&  \text{Tr}_B \left[ \hat{V}(t) \hat{\rho}_U(0) \hat{V}(t)^{-1} \right].
\eeqn
For the matrix element $\rho_S^{nm}(t)$ we obtain
\beq
\rho^{nm}_S(t) = \bra{n} \rho_S(t) \ket{m} =  \rho^{nm}_S (0) \ r_{nm}(t)
\eeq
where
\beqn
\nn r_{nm}(t) &=& \text{Tr}_B \left[ e^{ - \f{\omega_0 (n-m)}{2} \sum_i (\kappa_i(t) a_i^\dagger - \kappa_i^*(t) a_i)} \bigotimes_i \rho_{B_i} \right] \\
&=& \prod_i \text{Tr}_{B_i} \left[ e^{ - \f{\omega_0 (n-m)}{2} (\kappa_i(t) a_i^\dagger - \kappa_i^*(t) a_i)} \rho_{B_i} \right].
\eeqn
Here the difference with the case of a single spin-$1/2$ as the system is the appearance of the factor $- \f{\omega_0 (n-m)}{2}$ in the above expressions. One can make a closer analogy by  defining
\beq
\kappa_i^{nm}(t) = - \f{\omega_0 (n-m)}{2} \kappa_i (t)
\eeq
and writing $r_{nm}(t)$ in terms of $\kappa_i^{nm}(t)$ as follows
\beq
r_{nm}(t) = \prod_i \langle e^{ (\kappa^{nm}_i(t) a_i^\dagger - \kappa^{nm^*}_i(t) a_i)} \rangle_{\rho_{B_i}}
\eeq
As discussed in \cite{Schlosshauer2007} each term of the above product is just the symmetrically ordered characteristic function of a single harmonic oscillator in thermal equilibrium and is 
\beq
\langle e^{ (\kappa^{nm}_i(t) a_i^\dagger - \kappa^{nm^*}_i(t) a_i)} \rangle_{\rho_{B_i}} = e^{-\f{1}{2} \abs{\kappa^{nm}_i(t)}^2 \coth (\omega_i / 2 k_B T)} 
\eeq 
Therefore,
\beq
r_{nm}(t) = e^{- (n-m)^2 \Gamma(t)}  
\eeq
where
\beq \label{Gamma}
\Gamma (t) = \omega_0^2 \sum_i \f{\lambda_i^2}{\omega_i^2 }(1 - \cos \omega_i t) \coth (\omega_i/ 2 k_B T)  
\eeq
Note that $\Gamma(t)$ is dimensionless as expected. 

So, in sum, the reduced matrix element evolves as 
\beq \label{rhonmt}
\rho_S^{nm}(t) = \rho_S^{nm} (0) e^{- (n-m)^2 \Gamma(t)}
\eeq
where $\Gamma(t)$ is given by Eq. \eqref{Gamma}. 

The first consequence of \eqref{rhonmt} is that the diagonal matrix element in the energy basis of the oscillator system are preserved, 
\beq
\rho_S^{nn}(t) = \rho_S^{nn}(0)
\eeq
For the off-diagonal matrix elements let us study the finite bath and infinite bath seperately. 

\subsection{Finite Heat Bath and False Decoherence}
Suppose the bath is composed of finite number of oscillators, denoted by $N$. We want to study the system at $T=0$, where the finite oscillators are in their ground states. In this case, 
\beq \label{Gam}
\Gamma (t) = \omega_0^2 \sum_{i=1}^N \f{\lambda_i^2}{\omega_i^2 }(1 - \cos \omega_i t).   
\eeq
Initially $\Gamma(0) = 0$. Does $\Gamma(t)$ return to its initial value after some finite time ? 
Ref. \cite{Schlosshauer2007} answers this question in the affirmative for the spin-$1/2$ system. However, in our opinion, for both the spin-$1/2$ system discussed there and the oscillator system discussed here, the answer is {\it in general} negative unless the periods of the terms in \eqref{Gam} have a common multiple! 

The reason is as follows. $\Gamma(t)$ is a finite sum of singly periodic functions $f_i(t)$ with fundamental (or least) periods $T_1, \cdots , T_N$, 
\beq
\Gamma(t) = \sum_{i=1}^N f_i(t).
\eeq
Here $f_i(t) = \f{\omega_0^2}{\omega_i^2} \lambda_i^2 (1 - \cos \omega_i t)$ and $T_i = 2 \pi / \omega_i$. If $\Gamma(t)$ is periodic with period $T$, 
\beq \label{GammaT}
\Gamma(t + T) = \Gamma(t),
\eeq
in general $f_i(t)$ must have the same period, 
\beq
f_i(t + T) = f_i(t)
\eeq
in order to satisfy \eqref{GammaT}. However, any period of a singly periodic function is an integer multiple of its fundamental period. So, there exist positive integer numbers $n_1, \cdots, n_N$ such that 
\beq
T =  n_1 T_1 = \cdots = n_N T_N.
\eeq
Therefore, the periodicity of $\Gamma(t)$ implies that for every two fundamental frequencies of the finite bath one should have
\beq \label{ToT}
\f{T_i}{T_j} = \f{\omega_j}{\omega_i} = \f{n_j}{n_i} \quad \text{(condition of false decoherence)}
\eeq
for some \emph{integer} numbers $n_i$, $n_j$. However, if there are two fundamental frequencies whose ratio $\f{\omega_j}{\omega_i}$ is not a \emph{rational number} then $\Gamma(t)$ is not periodic! 

Conversely if the fundamental periods of $f_i(t)$ satisfy \eqref{ToT} then $\Gamma(t)$ is periodic with fundamental period equal to least common multiple of $T_1, \cdots, T_N$.
In this case the coherence of the system will be restored by time $T$. This is one type of a class of phenomena called {\it false decoherence} where the off diagonal matrix elements of the density matrix diminish at small times but after some large time are restored. To my knowledge, false decoherence cannot usually be realized if one uses the Born-Markov approximation. 

The other occasion that false decoherence may occur is in some of the adiabatic interactions where Born-Oppenheimer approximation is relevant. One can find an excellent discussion of  this point by Leggett in first two lectures of \cite{Leggett1989}

\begin{figure}[htbp]
\begin{center}
\includegraphics[scale=.6] {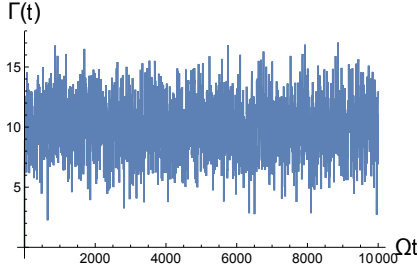}
\caption{Decoherence rate function $\Gamma(t)$ for finite oscillator bath with only 10 fundamental frequencies with irrational ratios. $\Gamma(t)$ starts at zero and increases to average value of about $10$. It fluctuates most of the times between $5-15$ and rarely reaches minimum values $2-3$, but it never returns to zero! Coherence is lost most of the times by at least by a factor of $e^{-5}$-$e^{-15}$. On the rare occasions when $\Gamma(t)$ becomes $\sim 2$ still $\sim 87 \%$  of the coherence is lost, $e^{-2} \sim 1 - 87\%$. }
\label{Gammat}
\end{center}
\end{figure}

If \eqref{ToT} is not satisfied for at least two fundamental frequencies of the bath, then one has {\emph true decoherence} and the system never returns to its initial coherent state. We demonstrate this fact below through an example. Consider a finite oscillator bath with $N=10$ and coupling coefficients and frequencies as follows
\beqn
\f{\omega_0^2 \lambda_n^2}{\omega_n^2} &=& 1, \\
\omega_n &=& \sqrt[3]{n} \ \Omega
\eeqn
for some $\Omega = const.$. The decoherence rate function becomes in this example 
\beq
\Gamma(t) = \sum_{n=1}^{10} (1 - \cos \sqrt[3]{n} \Omega t)
\eeq
Clearly not all pairs of frequencies satisfy \eqref{ToT}. So we expect the decoherence rate factor to increase and never returns to zero. This is indeed the case as Fig. \ref{Gammat} illustrates (see the caption). 

This may be considered as rather a new type of irreversible processes which have not received enough attention in the literature. Here there is no heat exchange between the bath and the system because the interaction Hamiltonian is diagonal in the energy basis of the system. Furthermore, the bath is finite. Nevertheless, the system does not return to its initial state and the decoherence is \emph{true}.  

\subsection{Infinite Bath}
For an infinite bath one may define a spectral density function as follows \cite{Blencowe2013},
\beq
J(\omega) = \pi \omega_0^2 \sum_i \lambda_i^2 \delta(\omega - \omega_i)
\eeq
Then $\Gamma(t)$ of Eq. \eqref{Gamma} can be written in terms of the spectral density function for an arbitrary temperature as 
\beq \label{Gint}
\Gamma(t) = \f{1}{\pi} \int J(\omega) \f{1- \cos \omega t}{\omega^2} \coth(\omega/2 k_B T) \ d\omega
\eeq
The integration is over the bandwidth that the frequencies of the bath oscillator make. We assume that the infinite oscillators make a continuum of frequencies and a smooth spectral density function $J(\omega)$. Further we assume that the frequencies have lower $\omega_L$ and upper $\Lambda$ cutoffs. For the ohmic spectral density of Ref. \cite{Blencowe2013} we adapt a sharp lower cutoff and a smooth upper cutoff as follows
\beq
J(\omega) = C \ \omega \ e^{-\omega/\Lambda} \ \theta (\omega- \omega_L)
\eeq
where $C$ is system-bath dimensionless coupling constant and $\theta(x)$ is the Heaviside step function.  Using the results of \cite{Schlosshauer2007} we take the integral (\ref{Gint}) for $k_B T \ll \Lambda$. The result is  
\beq
\Gamma(t) = \Gamma_{vac} (t) + \Gamma_{therm} (t)
\eeq
where 
\beq
\Gamma_{vac} (t)= \f{C}{2 \pi} \ln ( 1 + \Lambda^2 t^2)
\eeq
is the vacuum fluctuation decoherence rate function and 
\beq
\Gamma_{therm} (t) \approx \f{C} {\pi} \ln \left[ \f{\sinh(\pi k_B T t)}{\pi k_B T t} \right] - \f{C}{ \pi} k_B T \ \omega_L \ t^2
\eeq
is the thermal fluctuation decoherence rate function due to the bath. 

At $T=0$, the thermal fluctuation vanishes and the decoherence is entirely due to the vaccum
\beq
\Gamma(t) = \Gamma_{vac} (t)
\eeq
This is the case that cannot be obtained by use of Born-Markov approximation \cite{Blencowe2014pc}

For small and finite temperatures and short times, i.e. $k_B T t \ll 1$, the thermal fluctuation term becomes
\beq
\Gamma_{therm} \approx \f{1}{\pi} C k_B T (\f{\pi^2}{6} k_B T - \omega_L) t^2
\eeq

Finally for high temperatures and longer times where one has $\Lambda \gg k_B T \gg 1/t \gg \omega_L$ one recovers the Born-Markov approximation result to leading order,
\beq
\Gamma(t) \approx C k_B T t
\eeq
which agrees with Ref. \cite{Blencowe2013}. 

\subsection{Summary}
In summary we found almost exactly the decoherence rate of the toy model of gravitationally induced decoherence problem without use of Born-Markov approximation. We found that the off diagonal matrix elements decay as 
\beq 
\rho_S^{nm}(t) = \rho_S^{nm} (0) e^{- (n-m)^2 [\Gamma_{vac}(t) + \Gamma_{therm}(t)]}
\eeq
In the high temperature limit we recovered the result of Born-Markov approximation while at zero temperature we found that the thermal decoherence vanishes and the dominant decoherence is governed by the vacuum fluctuation. 

For the finite bath we found that decoherence takes place unless the ration of fundamental frequencies of the bath are rational numbers. 

\bibliographystyle{plain}
\bibliography{IOPEXPORT_BIB} 

\end{document}